# The influence of physical and mental constraints to a stream of people through a bottleneck

Paul Geoerg[1], Jette Schumann[2], Maik Boltes[2], Stefan Holl[2], Anja Hofmann[1]
[1]Bundesanstalt für Materialforschung und –prüfung, 12205 Berlin, Germany
paul.geoerg@bam.de; anja.hofmann@bam.de
[2]Forschungszentrum Jülich GmbH, 52425 Jülich, Germany
j.schumann@fz-juelich.de; m.boltes@fz-juelich.de; st.holl@fz-juelich.de

**Abstract** - Understanding movement in heterogeneous groups is important for a meaningful evaluation of evacuation prediction and for a proper design of buildings. The understanding of interactions and influencing factors in heterogeneous groups on key performance figures is fundamental for a safe design. This contribution presents results of experimental studies on movement of a crowd through a bottleneck involving participants with and without disabilities. High precise trajectories of the attendees extracted from video recordings were used to calculate density and velocity of the participants. Besides the well-established fundamental diagram new insights into the individual relation between density and velocity are discussed. A complex structure and considerate behaviour in movement implicates a strong influence of the heterogeneity on key performance values of safe movement.

## 1. Introduction

Many studies to investigate pedestrians' movement, their decision making, navigation and behaviour have been carried out during the last years [1-3]. Yet those studies mainly have analysed the movement of young and able-bodied participants in homogeneous groups. Thus, the studies do not represent circumstances in real life with a wide variety of individual limitations on moving. Empirical data sets involving people with disabilities are needed for a proper modelling and validation of heterogeneous pedestrian streams. Up to now only a few studies involving people with disabilities have been carried out and commonly focus on the unimpeded walking speed [4-8].

This contribution presents results of large-scale studies carried out in an interdisciplinary research-project "SiME - Safety for people with physical, mental or age-related disabilities" considering persons with body-, mental- or age-related disabilities as well as heterogeneous populations in evacuation planning. The experimental design and calculation methods are described in Sec. 2. The results will be discussed in Sec. 3 and lastly we conclude in Sec. 4.

## 2. Method

The characteristics of movement in heterogeneously composed populations are insufficiently described so far. Meanwhile, reduction of restrictions by disabilities is societal consensus and is connected to an old discussion about equal access vs equal egress [10]. The research project SiME focuses on safety in inclusive societies. As a part of this project, the relationship between inhomogeneity in groups and pedestrians' movement is studied. Twelve studies with more than 145 single runs and overall 252 participants with different difficulties and without any disabilities were performed in 2017 in a large hall in Wermelskirchen, Germany. One of these studies investigated the influence of body-related disabilities on the walking behaviour in groups.

## 2.1. Study setup

The main focus of this paper are two configurations comparing the movement of the crowd through a bottleneck with and without participating wheelchair users. The experimental setup is sketched in Fig. 1. Each run was carried out twice with the same boundary conditions and was captured by eight high definition cameras mounted at the ceiling of the hall more than 6 m above ground.

In the first study approx. 10 % were of the population as wheelchair users (82 participants without disabilities and seven wheelchair users at a mean age of $37 \pm 16$); the second study was performed without any participants with disabilities (69 participants at a mean age of $32 \pm 16$).

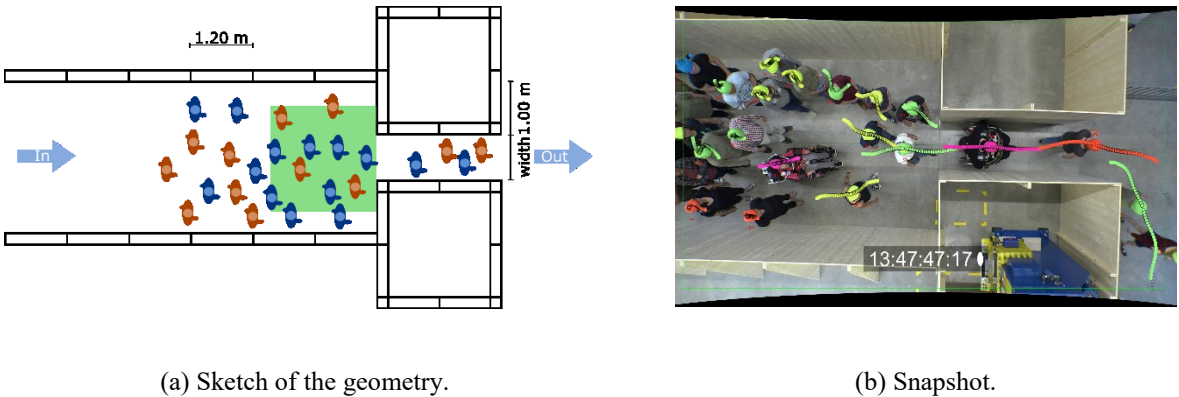

(a) Sketch of the geometry.            (b) Snapshot.

Fig 1: Geometrical setup of the study (left) and snapshot (right) of a study with wheelchair users and tracked individual positions for a bottleneck width of 0.9 m.

All attendees were asked to move directly and quick through the bottleneck but without pushing each other. Participation was voluntary for everybody and a cancellation of participation without any negative consequences at any time was possible. All participants have been paid for participation by 25, 00 € per half a day. Only anonymous data were used for the studies. The ethics committee of the University of Wuppertal has approved the project. No ethical concerns were mentioned.

## 2.2. Data extraction

The passageway through the geometry setup was captured by cameras and processed by a software developed for automatic extraction of attendees' trajectories, PeTrack [11]. For extracting individual positions at every frame, the positions of the coloured caps were detected and tracked (for example see Fig. 1(a) with the position of participants highlighted one second in the past and future). Participants were wearing coloured caps to encode their height. The centre of the coloured area auto corrected by taking the perspective view into account [12] results in trajectories for all participants. By neglecting the height information the trajectories represent the path on the ground.
We assume a coherent area of pixels in the middle of the coloured hats as body centre of a participant, so the resulting trajectories represent positions of the head projected on the ground. All resulting trajectories were checked and corrected manually. For more details we refer to [11].

## 2.2. Calculation methods

To quantify characteristics of participants' movement the Voronoi method is used to calculate density and velocity [9, 13]. Density and velocity were analysed in a rectangular measurement area of 2 x 2 m centred in front of the bottleneck entrance (green dotted area in Fig. 2). Calculations were utilised by *JPSreport* (part of the *JuPedSim*-Framework [14]).

## 3. Analysis and results

To compare attendees' movement considering participants with disabilities, we analysed movement data from studies with and without wheelchair users. Fig. 2 shows trajectories for two runs with a bottleneck width of 0.9 m. Based on these trajectories, movement characteristics such as density, velocity and flow at any time and position are calculated.

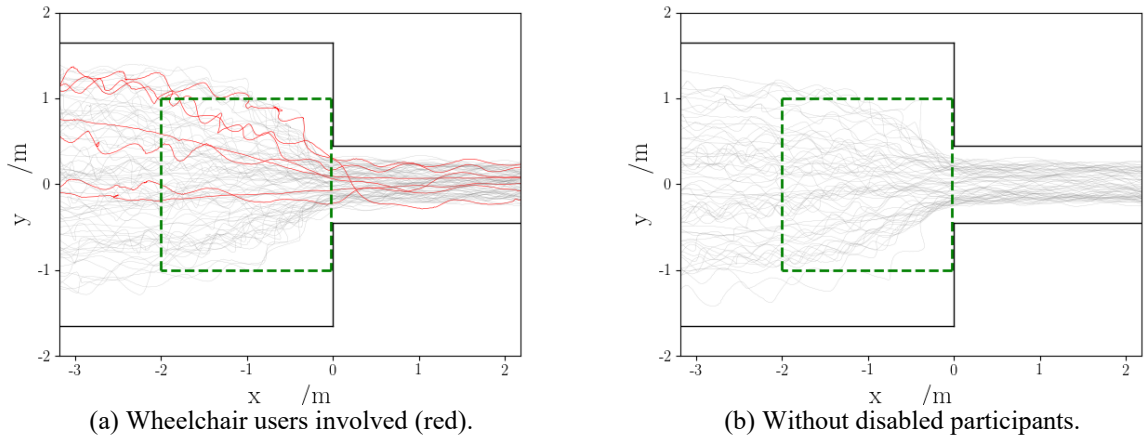

(a) Wheelchair users involved (red).  (b) Without disabled participants.

Fig 2: Attendees' trajectories from studies with wheelchair users (red) and without (grey) disabilities passing the bottleneck (width = 0.9 m). Measurement area for calculating speed and density is represented by the green dotted rectangle.

The time-dependent development of density and velocity is characterised by medium densities (mean density: with wheelchair users $1.50 \pm 0.72$ /m$^{-2}$, without wheelchair users: $1.93 \pm 0.94$ /m$^{-2}$) and significant slow individual velocities (mean velocity: with wheelchair users: $0.35 \pm 0.23$ m/s, without wheelchair users: $0.37 \pm 0.23$ m/s). Taking into account that the studies are conducted for short time periods a significant percentage of the run is sensitive from start and end conditions. It is common to use the steady part of a run for further analysis which is independent to start and end conditions [15, 16].

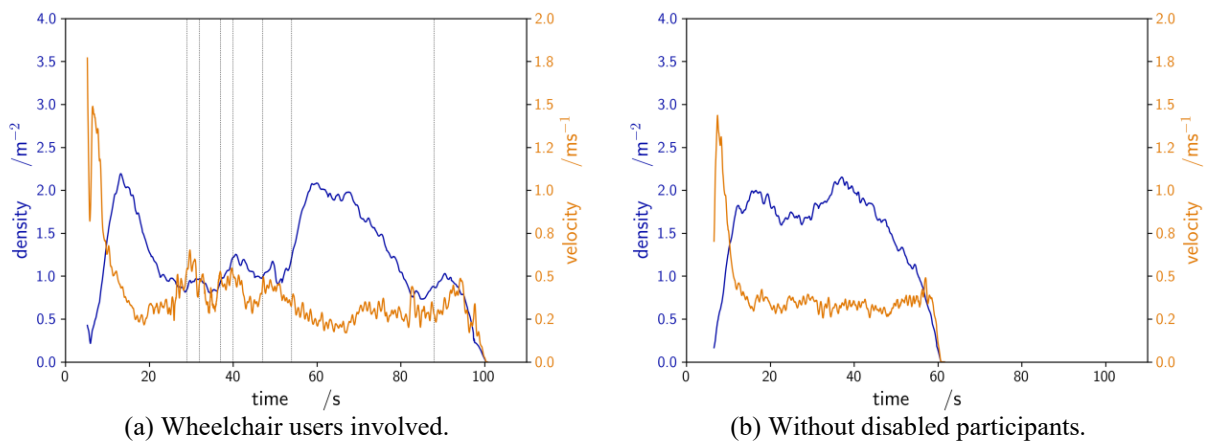

(a) Wheelchair users involved.  (b) Without disabled participants.

Fig 3: Time series for the development of density and velocity for two runs with and without wheelchair users. The time of entrance of a wheelchair user in the bottleneck is indicated by black dotted vertical lines. Steady states can be hardly defined by the relative stable stage of the curves.

For both runs the density increases at the beginning of the run and decreases at the end. Development of density in the study without wheelchair users fluctuates slightly in the middle of the run. The characteristic of the run considering wheelchair users is significantly different: density and velocity are characterised by strong fluctuations. In this case, density and velocity are influenced by other events than start or end conditions and boundaries. The presence of wheelchair users becomes important and causes significant fluctuations in time development of density and flow. Fluctuations occur when a wheelchair user reaches the entrance of the bottleneck (see Fig. 3(a)). Wheelchair users may cause a rapid decrease of density and velocity in the measurement area located in front of the bottleneck. To consider this striking fluctuations in the analysis (e.g. in density-speed-relations, see Fig. 5), data over the entire time-interval is considered.

The fluctuations observed in the time series are also exhibited in the representation of the trajectories in space and time (see Fig. 4). We can show, that participating wheelchair users tend to favour stop-and-go waves. To gain a deeper understanding the fundamental relationship between density and velocity is analysed (see Fig. 5).

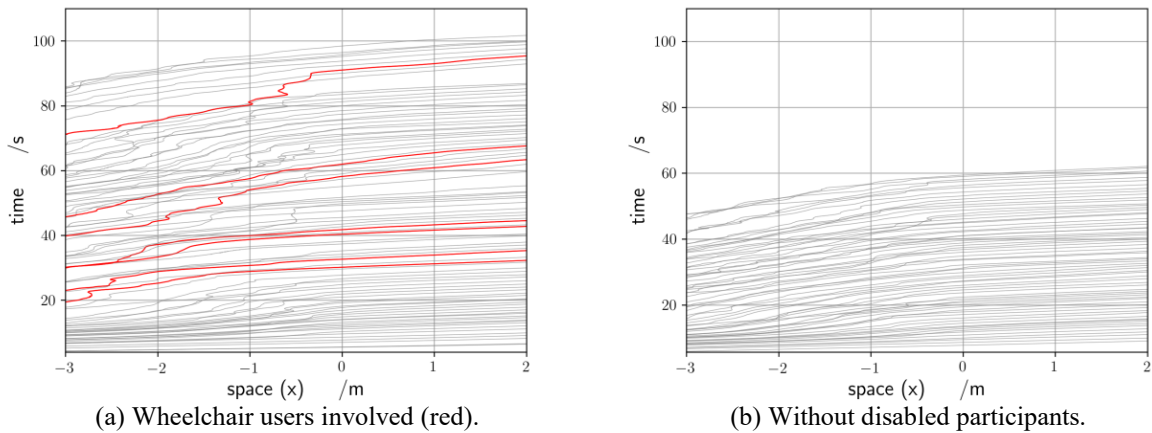

(a) Wheelchair users involved (red).      (b) Without disabled participants.

Fig 4: Time-space-relations for two runs with and without wheelchair users. Participating wheelchair user (red lines) lead to an occurrence of stop-and-go waves.

We found values for participants in wheelchairs in all areas of the speed-density-relation, especially in density-areas $\leq 3.0$ /m$^{-2}$. It is striking that density and velocity are not dependent. So surprisingly the expected three basic domains free flow, bound flow and congestion [17, 18] are not observed. A plateau of velocities is formed for densities above $\geq 1.0$ /m$^{-2}$. Such plateaus are known from ordering effects in bidirectional flows [19, 20]. Even if a wheelchair user reaches the bottleneck, their neighbours anticipate individual velocity and stop passing. Participants interact and solve priority of movement by communication.

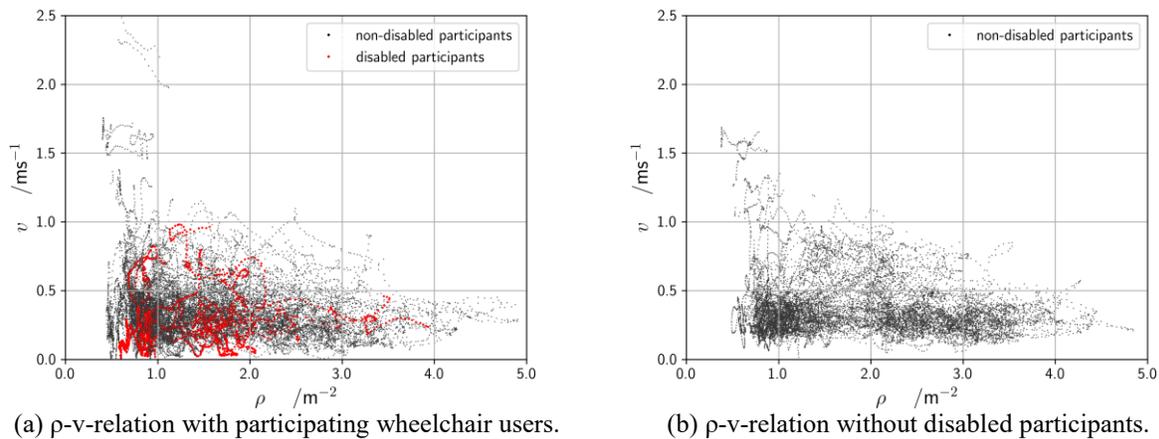

(a) ρ-v-relation with participating wheelchair users.   (b) ρ-v-relation without disabled participants.

Fig 5: Individual speed-density-relation for a run with wheelchair users. Individual densities and velocities for wheelchair users are coloured in red.

This behaviour can be interpreted as ordering effects expecting the bottleneck (see the increasing inhomogeneity prior to the bottleneck-entrance in Fig. 4). Social norms, individual behaviour and degrees of freedom in movement as well as the influence of technical assistance devices and accompanying persons may affect the passageway. Further on, the impact of disabilities on movement differs: those who are in a wheelchair may be able to roll with a high unimpeded movement speed, but may characterised by a delayed acceleration because of the inertia of the wheelchair.

## 3. Conclusions

A series of well-controlled laboratory studies on movement through a bottleneck participating wheelchair users and participants without disabilities was analysed. Movement of the crowd was recorded with high-definition cameras and precise trajectories of all participants were extracted automatically. Relations between individual density and velocity calculated with the Voronoi method were presented. Usually expected relation between density and velocity was not observed: the movement speed in heterogeneous groups depends not only on pedestrians' density but also on singular events like communication in front of the bottleneck, considerate behaviour or anticipation of social norms. This leads to significant fluctuations in speed and density and indicates the importance of social behaviour on the performance of a facility. The presented results contribute to empirical data of pedestrian dynamics and arise a lot of questions regarding the complexity of movement in heterogeneous social groups.

**Acknowledgements**

Part of this work has been performed within the research program Safety for people with physical, mental or age-related disabilities (SiME) supported by the German Federal Ministry of Education and Research - BMBF (FKZ starting on 13N13946). Paul Geoerg thanks the SFPE Foundation for financial support with the Dr. Guylène Proulx, OC Scholarship.